\documentclass[sn-basic,pdflatex]{sn-jnl}

\usepackage{graphicx}%
\usepackage{multirow}%
\usepackage{amsmath,amssymb,amsfonts}%
\usepackage{amsthm}%
\usepackage{mathrsfs}%
\usepackage[title]{appendix}%
\usepackage{xcolor}%
\usepackage{textcomp}%
\usepackage{manyfoot}%
\usepackage{booktabs}%
\usepackage{algorithm}%
\usepackage{algorithmicx}%
\usepackage{algpseudocode}%
\usepackage{listings}%
\theoremstyle{remark}
\theoremstyle{definition}

\usepackage{booktabs}
\usepackage{longtable}
\usepackage{array}
\usepackage{multirow}
\usepackage{wrapfig}
\usepackage{float}
\usepackage{colortbl}
\usepackage{pdflscape}
\usepackage{tabu}
\usepackage{threeparttable}
\usepackage{threeparttablex}
\usepackage[normalem]{ulem}
\usepackage{makecell}
\usepackage{xcolor}

\providecommand{\tightlist}{%
  \setlength{\itemsep}{0pt}\setlength{\parskip}{0pt}}

\begin{document}

\title[Resolving power]{Resolving power: A general approach to compare
the distinguishing ability of threshold-free evaluation metrics}

%%=============================================================%%
%% Prefix	-> \pfx{Dr}
%% GivenName	-> \fnm{Joergen W.}
%% Particle	-> \spfx{van der} -> surname prefix
%% FamilyName	-> \sur{Ploeg}
%% Suffix	-> \sfx{IV}
%% NatureName	-> \tanm{Poet Laureate} -> Title after name
%% Degrees	-> \dgr{MSc, PhD}
%% \author*[1,2]{\pfx{Dr} \fnm{Joergen W.} \spfx{van der} \sur{Ploeg} \sfx{IV} \tanm{Poet Laureate}
%%                 \dgr{MSc, PhD}}\email{iauthor@gmail.com}
%%=============================================================%%

\author*[]{\fnm{Colin} \sur{Beam} \dgr{PhD}}\email{\href{mailto:colinbeam@gmail.com}{\nolinkurl{colinbeam@gmail.com}}}

  \affil*[]{\orgname{UCLA Health}, \orgaddress{\city{San Luis
Obispo}, \country{United States}}}

\abstract{Selecting an evaluation metric is fundamental to model
development, but uncertainty remains about when certain metrics are
preferable and why. This paper introduces the concept of \emph{resolving
power} to describe the ability of an evaluation metric to distinguish
between binary classifiers of similar quality. This ability depends on
two attributes: 1. The metric's response to improvements in classifier
quality (its signal), and 2. The metric's sampling variability (its
noise). The paper defines resolving power generically as a metric's
sampling uncertainty scaled by its signal. A simulation study compares
the area under the receiver operating characteristic curve (AUROC) and
the and the area under the precision-recall curve (AUPRC) in a variety
of contexts. It finds that the AUROC generally has greater resolving
power, but that the AUPRC is better when searching among high-quality
classifiers applied to low prevalence outcomes. The paper also proposes
an empirical method to estimate resolving power that can be applied to
any dataset and any initial classification model. The AUROC is useful
for developing the resolving power concept, but it has been criticized
for being misleading. Newer metrics developed to address its
interpretative issues can be easily incorporated into the resolving
power framework. The best metrics for model search will be both
interpretable and high in resolving power. Sometimes these objectives
will conflict and how to address this tension remains an open question.}


\maketitle

\section{Introduction}\label{sec1}

There is a large and growing collection of evaluation metrics used for
binary classification models. Choosing a metric can be challenging as
model evaluation serves a variety of goals. One is interpretation,
meaning that the metric is both easily understood and sensitive to
aspects of quality that are relevant to the user. Simple classification
accuracy, for example, is misleading for outcomes that are both low
prevalence and relatively severe, such as rare but serious diseases.
Evaluation metrics are also used to select the best model from a
collection of competitors \citep{raschka2018model}. This includes
selection between different model classes, such as between a simple
baseline model and more complex machine learning models. And it includes
selection within a model class, as occurs with hyperparameter search
during model tuning. Another goal of evaluation is to estimate how well
a given model will perform on future, unseen cases
\citep{saito2015precision}.

The Receiver Operating Characteristic (ROC) curve has become a favored
method for evaluating binary classification models, in part due to the
shortcomings of simple classification accuracy
\citep{fawcett2006introduction}. More recently, many have argued that
the precision-recall curve (PRC) is preferable when there is a strong
class imbalance and where there is low value in true-negative
predictions
\citep{boyd2013area, saito2015precision, davis2006relationship}.
Relative to the ROC curve, the PRC gives more weight to the
highest-ranked cases located in the ``early retrieval area'' of ROC
space. These cases are especially important when capacity to act is
limited, such as when a health system has resources to intervene on only
their sickest patients.

Sampling uncertainty is a neglected aspect of model evaluation within
the field of machine learning \citep{vabalas2019machine}, but it is
essential to account for when data is limited
\citep{boyd2013area, dietterich1998approximate}. There has been scant
research that compares the sampling precision of evaluation metrics for
binary classifiers. One exception is found in
\citet{zhou2023discriminating}, who used a link prediction task on a toy
network model with a tunable noise parameter. For these network models
Zhou finds that the area under the ROC curve (AUROC) and the area under
the PRC (AUPRC) are much more discriminating than ``balanced
precision'', and that the AUROC is slightly more discriminating than
AUPRC.\footnote{Zhou finds balanced precision by choosing the decision
  threshold so that precision equals recall.}

This paper pursues a general approach for comparing evaluation metrics.
It also seeks specific conclusions about when and by how much some
metrics are better than others. This project is conceptually difficult
since evaluation metrics themselves are used to measure quality, each
encoding different assumptions about what makes a model better or worse.
The paper's strategy is to use a collection of sampling models to
construct a quality dimension that serves as the common standard of
comparison. The sampling models are used to assess how an evaluation
metric responds to changes in model quality (its signal) and how much
variability it has at a given level of quality (its noise). These two
quantities are combined to form an evaluation metric's \emph{resolving
power}, which is a type of signal-to-noise ratio. The resolving power of
a microscope is its capacity to distinguish between two close objects.
By analogy, the resolving power of an evaluation metric describes how
well it differentiates between models of similar quality. More
specifically, resolving power is defined as a metric's sampling
uncertainty mapped to a common scale.

Resolving power draws inspiration from several previous works such as
\citet{saito2015precision}, \citet{Mazzanti2020}, and
\citet{hand2023notes}. Each of these analyses addresses a metric's
adequacy in describing performance. In contrast, this paper's focus on
metric sampling uncertainty primarily pertains to model search and
selection. The remainder of this paper presents the resolving power
methodology, demonstrates its application to the AUROC and the AUPRC,
and then reflects on the implications for model search.

\section{ROC and PR curves}\label{sec2}

Our interest is in models that map cases to predicted classes. A
\emph{discrete} classifier is one that only outputs a class label.
Applying a discrete classifier to test data produces a 2x2 confusion
matrix (Table \ref{tab:confusion_matrix}), with rows corresponding to
the predicted class and columns giving the true class. A \emph{scoring}
classifier outputs a number on a continuous scale, such as an estimated
probability, that represents the degree to which a case belongs to a
class \citep{fawcett2006introduction}. Applying a decision threshold to
a scoring classifier produces a discrete classifier.
\citet{hand2009measuring} shows that choosing a particular threshold is
equivalent to specifying the relative costs of false positives versus
false negatives.

\begin{table}[!h]
\centering
\caption{\label{tab:confusion_matrix}Example confusion matrix}
\centering
\begin{tabular}[t]{l|c|c}
\hline
  & actual + & actual -\\
\hline
predicted + & TP & FP\\
\hline
predicted - & FN & TN\\
\hline
total & P & N\\
\hline
\end{tabular}
\end{table}

A variety of familiar evaluation metrics may be calculated for discrete
classifiers such as accuracy, recall (hit rate, sensitivity, true
positive rate), precision (positive predicted value), specificity, and
the F1-score. These are known as single-threshold (or
threshold-dependent) metrics. In contrast, threshold-free metrics use
the full range of the original scores. Examples include the AUROC, the
AUPRC, and the area under the precision-recall-gain curve (AUPRG), among
others. Threshold-free metrics are advantageous since they allow users
to adapt the model to a specific context \citep{flach2015precision}. The
AUROC and AUPRC are preferred metrics when the primary goal is to
achieve good discrimination so that cases are efficiently sorted into
the positive and negative classes.

The ROC curve depicts the trade-off between the true positive rate (tpr)
on the y-axis and the false positive rate (fpr) on the x-axis. A
discrete classifier only gives a single point in ROC space,
corresponding to its one confusion matrix. A scoring classifier gives
points for every possible confusion matrix that can be formed by varying
the decision threshold. The empirical ROC curve interpolates between
these points to create a step function. As the number of points become
arbitrarily large the empirical curve will approach the population ROC
curve.

If a decision threshold is selected to flag 50 percent of all cases and
the classifier is no better than random guessing then we expect it to
identify half of the positives and half of the negatives, yielding the
point \((0.5, 0.5)\) in ROC space. Similarly, a random classifier
flagging 20 percent of cases is expected to have a recall of 20 percent
and a false positive rate of 20 percent. The random guessing classifier,
then, is given by the \(y = x\) line in ROC space. A perfect classifier
ranks all positive cases above all negative cases, so it corresponds to
the step function from \((0,0)\) to \((0,1)\) for all the positives, and
then from \((0,1)\) to \((1,1)\) for all the negatives. Classifiers that
lie above the identity line but below the perfect step function
represent intermediate performance with better classifiers containing
points closer to the \((0,1)\) northwest corner of ROC space.

The AUROC summarizes a classifier's performance across all decision
thresholds and is found by integrating the ROC curve over the \([0,1]\)
range of false positive rates. Larger AUROC values are better, with the
random classifier giving an \(\text{AUROC} = 0.5\) and the perfect
classifier giving an \(\text{AUROC} = 1\). The AUROC, as a scalar value,
is especially relevant for model tuning and selection. A disadvantage of
the AUROC is that it can conceal local differences in performance. For
instance, one classifier may be better for highly ranked cases while
another is better for those in the intermediate or lower ranks. An
important statistical property of the AUROC is that its value equals the
probability that a classifier will rank a randomly chosen positive case
higher than a randomly chosen negative case
\citep{green1966signal, hanley1982meaning}.

The AUROC can be interpreted as an average sensitivity, assuming all
specificity values are equally likely \citep{hand2009measuring}. Several
authors argue that the AUROC is deficient in key respects.
\citet{swamidass2010croc} remark that treating all specificity values as
uniformly important is not appropriate for most problems and propose the
area under the concentrated ROC as a corrective. \citet{byrne2016note}
shows that the AUROC is typically not a proper scoring function. The
AUROC has also been criticized as incoherent since two classifiers with
the same AUROC will typically imply different relative costs of
misclassification \citep[@hand2023notes]{hand2009measuring}. Hand
proposes the H-measure as a coherent alternative (Online Resource 1
applies resolving power to the H-measure). In response to Hand,
\citet{ferri2011coherent} argue for alternative interpretations of the
AUROC that are both coherent and model independent.

Precision-recall (PR) graphs plot precision on the y-axis and recall on
the x-axis. In PR space a random classifier corresponds to the
horizontal line \(y = \frac{P}{P+N}  = \text{prevalence}\) where \(P\)
is the number of positive cases and \(N\) is the number of negative
cases. PR curves are sensitive to class skew (meaning one class occurs
more than the other) while ROC curves are not. This is because inputs to
the ROC curve, the true and false positive rates, only depend on the
column sums of the confusion matrix. Precision depends on the row sum of
true and false positives, so all else equal, it will decrease with
decreasing prevalence. Insensitivity to skew has been described as both
an advantage \citep{fawcett2006introduction} and disadvantage
\citep{saito2015precision} of the ROC curve.

Just like the AUROC, the AUPRC reduces a scoring classifier's
performance to a single value, with larger values indicating better
performance. Similar to the AUROC, the AUPRC can be interpreted as the
classifier's average precision over the \([0,1]\) range of recall
values. \citet{davis2006relationship} demonstrate that there is a
one-to-one correspondence between empirical ROC and PR curves since they
both chart a unique mapping from confusion matrices to points in ROC or
PR space. They go on to show that the AUROC and AUPRC give the same
model rankings when one model's curve ``dominates'' another's.
Informally, one curve dominates another if it lies above or equal to it
across their domains. A dominating ROC curve will be northwest of the
dominated curve, where its tpr is higher, its fpr is lower, or both. And
a dominating PR curve will be northeast of a dominated curve, with
higher precision, recall, or both across the entire domain. When there
is no domination (when two curves cross) the AUROC and AUPRC can give
different rankings. In cases of disagreement, the AUPRC favors
classifiers with better performance in the early retrieval area, which
is the region of low false positive rates in ROC space.

Because it gives more weight to the early retrieval area, the
precision-recall curve is often recommended for highly-skewed datasets.
Yet the empirical PRC is an imprecise estimate of the true curve,
especially for small sample sizes and with strong class imbalance
\citep{brodersen2010binormal}. This raises the question of whether the
advantages of the PRC are worth its cost in precision. Answering this
question requires that we compare metrics measured on different scales.

\section{Mapping between metrics}\label{sec3}

ROC analysis was initially developed to evaluate electronic sensors,
such as radar, during World War II. In the 1950s research psychologists
elaborated ROC analysis under the rubric of signal detection theory
(SDT), which soon became influential within experimental psychology,
psychophysics, and cognitive neuroscience \citep{wixted2020forgotten}.
Fundamental to SDT is the specification of two probability
distributions: A noise distribution for trials when the signal is absent
and a signal distribution for trials when the signal is present
\citep{green1966signal}. The binormal model (two Gaussians) is the most
common choice for the signal and noise distributions. The SDT framework
can be described in the language of binary classification with signal
and noise trials considered members of the positive and negative
classes, respectively.

A classification model applied to feature measurements generates class
score distributions. For a simple example, suppose the two classes are
women and men and that there is one feature of height. The
classification model will just be the identity mapping applied to the
height measurements. The binormal model should then be a good
approximation for the score distributions.\footnote{Height is believed
  to result from the sum of a large number of independent genetic and
  environmental effects, so by The Central Limit Theorem the
  distributions should be approximately normal.} Figure
\ref{fig:binormal_example} shows the binormal model for this example,
using height distribution parameters from Our World in Data
\citep{roser2013human}. Women have an average height of 164.7 cm with a
standard deviation of 7.1 cm, while men have an average height and
standard deviation of 178.4 cm and 7.6 cm, respectively. The vertical
dashed line is an example of a decision threshold, where any person
above 171 cm is classified as a man and any below as a woman (this type
of rule might be used in low visibility contexts where height is the
most salient feature). Hit rates and false alarm rates can be calculated
for that decision threshold, giving one point in ROC space.

\begin{figure}
\includegraphics{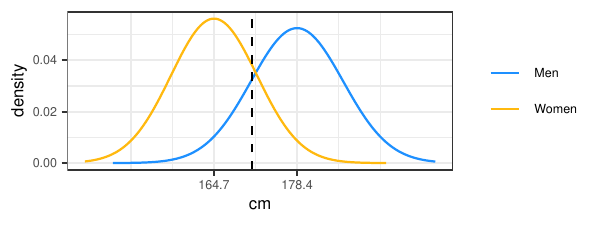} \caption{\label{fig:binormal_example}A binormal classifier example. The distribution of men's and women's heights approximately follow a normal distribution. The model implies an AUROC of .906. The vertical dashed line at 171 cm is an example decision threshold.}\label{fig:binormal_example}
\end{figure}

Now we must address what we mean by ``model quality''. This paper adopts
a discriminative conception: Better classifiers yield greater separation
in the class scores \citep{hand2001simple}. Importantly, model quality
refers to the out-of-sample class score distributions, which are
typically estimated with resampling methods or by using a test set. A
perfect classifier completely separates the class scores while a random
classifier has identical class score distributions.\footnote{\citet{hand2023notes}
  formally define a random classifier as one with identical cumulative
  distribution functions (cdfs) so that \(F_0 \equiv F_1\), where
  \(F_0\) and \(F_1\) are the cdfs of the negative and positive class
  scores, respectively. We can also use cdfs to define a perfect
  classifier: One where there exists a decision threshold \(t^*\) such
  that \(F_0(t^*) = 1\) and \(F_1(t^*)=0\). Finally, if the cdfs are not
  identical and they do not completely separate the class scores then
  they partially separate the distributions. That is, there is partial
  separation if \(F_0 \not\equiv F_1\) and there is at least one
  threshold \(t'\) such that \(F_0(t') < 1\) and \(F_1(t') > 0\).}
Classifiers that are neither random nor perfect are those that partially
separate the score distributions.

For the binormal model there is an analytic expression for the AUROC as
a function of the means and variances of the score distributions
\citep{marzban2004roc}. The binormal model parameters shown in Figure
\ref{fig:binormal_example} imply an \(\text{AUROC} =\) 0.906. Remember,
this implies there is about a 90 percent chance that a randomly chosen
man will be taller than a randomly chosen woman. In contrast to the
AUROC, the AUPRC is a function of both the score distributions and the
outcome prevalence. \citet{brodersen2010binormal} show how to
approximate the AUPRC for a given binormal model using numerical
integration.

One way to represent classifier improvement, such as occurs during model
tuning, is as a process that diminishes the overlap (increases the
separation) in the class scores. An ordered sequence of increasingly
separated distributions constructs a quality dimension that can unify
disparate metrics. For each set of distributions in the sequence we can
find the associated pairs of metric values. This forms a mapping that
can be used to compare metrics. The applications below use this approach
to trace a curve in the \(\text{AUROC} \times \text{AUPRC}\) plane.

The mapping between metrics depends on how the score distributions are
separated and we cannot know in advance how classifier improvement will
change the risk scores. There are myriad ways to increase class
separation, each indicative of different types of improvement. This
paper solves this ambiguity by fiat: It assumes that simple
manipulations of score distributions are a reasonable description of
model improvement. A simple approach, used below, is to add fixed
increments to the positive class scores. A concern is that a simple
additive mechanism may poorly describe how improvement happens in
practice. A more realistic, though more involved strategy is to base the
mechanism on the observed early stages of model improvement (Online
Resource 1 sketches this approach).

More generally, resolving power depends on two key assumptions: 1. That
model improvement is a mostly homogeneous process, and 2. That we can
approximate this process. The first assumption holds that improvement
largely occurs in similar ways across different algorithms or
hyperparameter settings. Though it does allow for random deviations due
to sampling or to the intrinsic stochasticity of some algorithms. The
second assumption is that we can do a decent job describing the common
improvement process by finding the right sequence of class score
distributions.

We have identified the quality dimension as an ordered sequence of
distributions, but how should we measure location on this dimension? One
option is to just use the model rankings themselves, which forms an
ordinal scale \citep{stevens1946theory}. Another option is to measure
distribution overlap directly using the Bhattacharyya coefficient. Or we
can use a measure that relates overlap to model quality, the AUROC and
AUPRC being two examples among many. The AUROC has several properties
that make it a good choice. Since the AUROC is an area (and a
probability), equal differences across the scale represent equal
differences in amount. Another advantage, mentioned above, is that it is
unaffected by outcome prevalence. The AUROC is also the most popular
threshold-free evaluation metric for binary classifiers, making it a
natural choice for the reference metric.

For our purposes, the AUROC's biggest advantage is that it is agnostic
with respect to where changes occur in the score distributions. This
fact is easiest to demonstrate with an empirical score distribution,
defined as a finite set of risk scores and associated outcomes. Briefly,
suppose there are \(n^+\) positive cases, \(n^-\) negative cases, and
that all risk scores are unique. Further suppose that the classifier is
not perfect, so \(0.5 \leq \text{AUROC} < 1\), and we want to improve
this by perturbing the risk scores. If we sort all cases together into a
single list ranked by score, then the smallest improvements occur by
finding pairs of adjacent scores that are ``out-of-order'', such that
the negative case has a higher score than the positive case, and
re-ordering these pairs. Re-ordering a single pair will improve the
AUROC by \(\frac{1}{n^+}\times\frac{1}{n^-}\) regardless of where the
improvement occurs. This follows from the probabilistic interpretation
of the AUROC: There are \(n^+ \times n^-\) unique ordered pairs of
positive and negative scores, which forms the number of events in the
sample space. So resolving one out-of-order pair increases the
probability by \(\frac{1}{n^+\times n^-}\). In contrast, the AUPRC will
improve more for resolving out-of-order pairs that are among the
highest-ranked risk scores.

To summarize this section's key points: Classifier quality is gauged by
its outputs, the class score distributions. A sequence of increasingly
separated class distributions forms a common quality dimension that
charts the relationship between different evaluation metrics. A key
caveat is that the mapping between metrics is contingent on how the
score distributions are separated. Several characteristics of the AUROC
make it a good choice as the reference measure of model
quality.\footnote{Note that these advantages pertain narrowly to the
  AUROC's role as reference measure and do not counter to the criticisms
  of the AUROC referenced in the previous section.} In particular, the
AUROC always improves by a constant amount when resolving a pair of
adjacent out-of-order risk scores. Moreover, the viability of resolving
power does not hinge on using the AUROC as the reference. Instead, it is
the sequence of class score distributions that are fundamental. Other
metrics can be substituted so long as they form a mapping from the class
score distributions to an interval scale.

\section{Resolving power}\label{sec4}

The resolving power method can be summarized in four steps:

\begin{enumerate}
\def\labelenumi{\arabic{enumi}.}
\tightlist
\item
  \textbf{Sampling model}: Specify class score distributions,
  prevalence, and sample size.
\item
  \textbf{Signal curves}: Use the sampling model to create a fine grid
  of improving classifiers. Find each metric's values across the grid.
\item
  \textbf{Noise distributions}: Estimate metric sampling uncertainty by
  drawing random samples at points of interest within the quality grid.
\item
  \textbf{Comparison}: Use the above results to estimate and compare
  resolving power.
\end{enumerate}

This section illustrates the core mechanics of the approach using an
idealized example while later sections move to the applications. Suppose
we are interested in comparing the resolving power of the AUROC with
that of the fictional ``area under the super-great curve'' (the AUSGC).
We construct a sampling model by specifying the class score
distributions, prevalence, and sample size. Next, we create a grid of
1000 models. The first model has identical class score distributions for
the random classifier and then we gradually shift the distributions
apart so that the 1000th model has almost no overlap. Finally, we want
to assess sampling models that give AUROC values of 0.7 and 0.9. We draw
many replicates from these two models to estimate the sampling
variability of the two metrics. Figure \ref{fig:toy_example} summarizes
the analysis.

\begin{figure}
\includegraphics{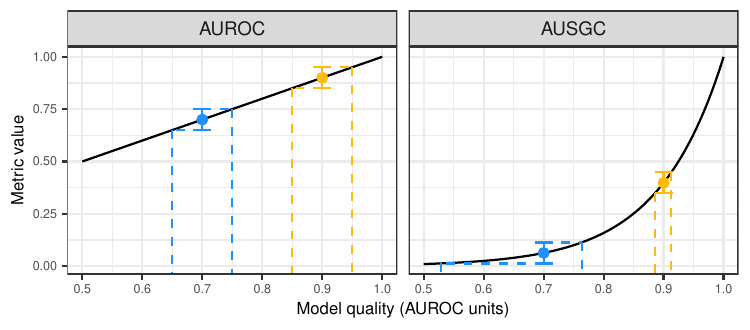} \caption{\label{fig:toy_example}Signal curve example. The two panels are united by the same sequence of models used to construct the quality grid. The AUROC serves as the common reference scale on the x-axis.}\label{fig:toy_example}
\end{figure}

Following the previous section's recommendation, the AUROC on the x-axis
serves as the reference scale for model quality. The left panel is then
just the identity mapping. The right panel shows how the AUSGC changes
relative to the AUROC, giving the relative signal of the two metrics
across the quality continuum. Unit slope indicates equal signal, a slope
less than 1 favors the AUROC, and a slope greater than 1 favors the
AUSGC. For a given point on the curve, repeated draws from the sampling
model estimate each metric's noise distribution. The signal curves allow
us to map each metric's uncertainty interval to a model quality
interval, which forms the common basis for comparison.

Previously, resolving power was defined generically as a metric's scaled
sampling uncertainty, but now we need to make this specific. Define a
metric's \emph{resolution} as the width of its 95 percent confidence
interval mapped to the quality scale. We denote this quantity with the
Greek letter \(\kappa\). A microscope's resolution limit is the smallest
distance between two points that can still be distinguished as separate
entities. Analogously, \(\kappa\) is the minimum distance for
statistical discrimination using the \(\alpha = .05\) convention from
null hypothesis testing. Resolving power is \(1/\kappa\), or just the
reciprocal of the resolution distance. With AUROC as the reference scale
we can form the following heuristic assessments: A resolving power of 10
is rather poor, 100 is decent, and 1000 is good. Of course, these
assessments will depend on the context. A resolving power of 100 is less
impressive for a sample size of one million than for one of ten
thousand.

A disadvantage of the resolving power definition is that it requires
choosing an arbitrary \(\alpha\) level. Appendix A describes an
alternative approach that eliminates this requirement by expressing
resolving power as a scaled standard error. This comes at a cost of
stronger assumptions: The alternate approach assumes that the signal
curve is locally well-approximated by a straight line and that the
evaluation metric's sampling distribution is roughly symmetric.

Returning to the example, for the AUROC 0.7 model shown in blue we have:

\begin{itemize}
\tightlist
\item
  An AUROC of .7 with a 95\% confidence interval of {[}.65, .75{]}.
\item
  An AUSGC of .063 with a 95\% confidence interval of {[}.013, .114{]}.
  This maps to an AUROC interval of {[}.53, .76{]}.
\end{itemize}

The dashed lines in Fig. \ref{fig:toy_example} show how the signal
curves map the confidence limits to a common quality scale. This is
trivial for the AUROC since it is the identity mapping. For the AUROC =
0.7 sampling model we can conclude that the AUSGC is much less precise
with a resolution of \(\kappa_{\text{SGC}} = .23\) compared to
\(\kappa_{\text{ROC}} = .1\) for the AUROC. Turning to the AUROC = 0.9
model shown in orange, we have:

\begin{itemize}
\tightlist
\item
  An AUROC of .9 with a 95\% confidence interval of {[}.85, .95{]}
\item
  An AUSGC of .4 with a 95\% confidence interval of {[}.35, .45{]}. This
  maps to an AUROC interval of {[}.89, .91{]}.
\end{itemize}

Note that the confidence intervals in the original metrics have stayed
the same width at .1 for both the AUROC and the AUSGC. However, the
AUSGC is now in a steeper region of the curve, so its signal-to-noise
ratio has improved. As a result, we obtain
\(\kappa_{\text{SGC}} = .02\), giving the AUSGC much better metric
resolution. From this analysis we can conclude that the AUSGC is only
``super-great'' when the search space spans a region of high-quality
models.

\section{Binormal model}\label{sec5}

The binormal model, as the most commonly used in ROC analysis, serves as
a good initial application of the approach. All code and data used below
are available on GitHub.\footnote{\url{https://github.com/colinbeam/resolving_power}}
Now we apply the four steps of the resolving power method.

\textbf{Step 1: The sampling model.} Assume a binormal model where
negative class scores have a standard normal \(\mathcal{N}(0,1)\)
distribution and positive class scores have \(\mathcal{N}(\delta_i,1)\)
distributions. The analysis explores a range of prevalence comprising
the values {[}.01, .05., .10, .20, .30, .40, .50{]}, which is the same
set used by \citet{Mazzanti2020}. We explore a moderately sized
classification task of 10,000 instances, so the lowest prevalence
condition has 100 instances in the positive class.

\textbf{Step 2: Signal curves.} Create a fine grid of improving models
by increasing the distance \(\delta_i\) between distributions. The grid
begins with the random classifier \(\text{AUROC}_1 = .5\) and ranges to
a max \(\text{AUROC}_n = .99995\). Each \(\delta_i\) is chosen to create
\(.00005\) AUROC increments between grid points. Since we have fixed
three of the four binormal model parameters, we can find the shift
parameter \(\delta_i\) as a function of the target \(\text{AUROC}_i\)
value (see \citet{marzban2004roc} for details). \[
\tag{1}
\delta_i = \sqrt{2}\times\Phi^{-1}(\text{AUROC}_i)
\] where \(\Phi^{-1}\) is the inverse cumulative standard normal
distribution. Note that an evenly spaced AUROC grid will require
progressively larger shifts between class distributions as model quality
increases. Next, we need to find the AUPRC values associated with each
AUROC grid point. The AUPRC can be found from a binormal model via
numerical approximation. Let \(\alpha\) represent the outcome prevalence
and \(\Phi_+\) and \(\Phi_-\) represent the cumulative Gaussian
distributions for the positive and negative classes, respectively.
\citet{brodersen2010binormal} derive the PR curve by finding precision
(PPV) as a function of recall (TPR): \[
\tag{2}
\text{PPV} = \frac{\alpha \text{TPR}}{\alpha \text{TPR} + (1-\alpha)\left(1 - \Phi_- \left(\Phi_+^{-1}(1 - \text{TPR}) \right) \right)} 
\] And to find the AUPRC they numerically approximate the integral: \[
\tag{3}
\text{AUPRC} = \int_0^1 \text{PPV}(\text{TPR})d\text{TPR}
\]

To summarize the steps: First we create an evenly spaced grid of AUROC
values using the implied shift parameter values from equation (1). We
then use the shift values in equation (2), specifically for the
\(\Phi_+\) parameterization. This gives us the PR curve so that we may
use equation (3) to find the associated AUPRC value.

Figure \ref{fig:gauss_SC} shows the binormal signal curves for each
condition. The relationship between metrics becomes more curvilinear as
prevalence decreases. This implies that, all else equal, the AUPRC will
be relatively more discriminating among higher quality models applied
within low prevalence contexts. The signal curve for a prevalence of .5
is approximately a straight line with an intercept of zero and a slope
of one---the identity mapping. Thus, the AUPRC and AUROC are estimating
the same quantity but by using different formulas. In this condition,
then, differences in resolving power will be due to differences from
sampling error alone.

\begin{figure}
\includegraphics{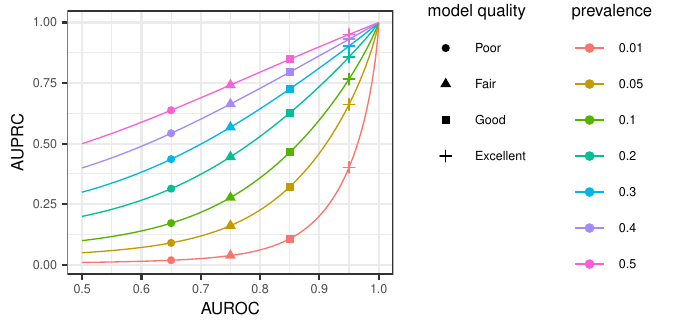} \caption{\label{fig:gauss_SC}Mapping between AUROC and AUPRC for the binormal model.}\label{fig:gauss_SC}
\end{figure}

\textbf{Step 3: Noise distributions.} We wish to assess a range of
quality values by evaluating models with AUROCs of
\([.65, .75, .85, .95]\). In the figures below these models are
respectively labeled ``Poor'', ``Fair'', ``Good'', and ``Excellent''.
For the four points of model quality we take 10,000 random samples from
each implied binormal model and estimate AUROC and AUPRC values with the
\texttt{PRROC} R package \citep{grau2015prroc}. The AUPRC is estimated
using the Davis and Goadrich method \citep{davis2006relationship}.
Ninety-five percent confidence intervals are found from the .025 and
.975 quantile values of the simulation samples.

\textbf{Step 4: Comparison.} For the final step we use the curves in
Figure \ref{fig:gauss_SC} to map the AUPRC 95 percent confidence
interval to the AUROC scale. We then find the relative difference in
metric resolution with the AUROC as the baseline (equal to the relative
difference in resolving power with the AUPRC as baseline): \[ 
\Delta = \frac{\kappa_\text{PRC} - \kappa_\text{ROC}}{\kappa_\text{ROC}} = \frac{1/\kappa_\text{ROC} - 1/\kappa_\text{PRC}}{1/\kappa_\text{PRC}} 
\] The simulation was repeated three times and estimates were averaged
to smooth out their variability across runs.

\begin{figure}
\includegraphics{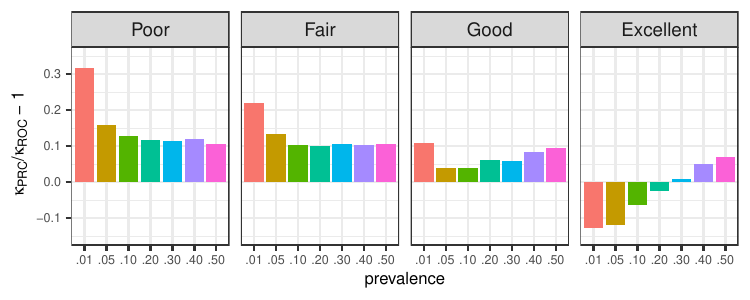} \caption{\label{fig:N10k_compare}Relative metric resolution by outcome prevalence and model quality for a binormal model with a sample size of N = 10,000. At each level of model quality 10,000 simulations are taken from the sampling model. Confidence limits are the .025 and .975 quantile values of the simulation samples.}\label{fig:N10k_compare}
\end{figure}

Simulation results are shown in Figure \ref{fig:N10k_compare}. Beginning
with the prevalence = .5 ``identity mapping'' condition, we see that the
AUPRC is usually around 10 percent more variable than the AUROC, though
the disadvantage is smaller in the ``Excellent'' model condition. In the
remaining conditions the AUPRC suffers a greater disadvantage in the
flatter portions of the signal curves, corresponding to contexts of low
prevalence and poor model quality. Specifically, the AUPRC is at a
disadvantage for all poor (AUROC = .65), fair (AUROC = .75), and good
(AUROC = .85) models across all levels of prevalence. AUPRC resolution
is typically about 10 percent larger, though in the flattest portion of
the signal curve---the low prevalence and low model quality
condition---the disadvantage is around 30 percent. The AUPRC has better
resolving power only for excellent models (AUROC = .95) applied to
moderately to strongly skewed datasets (prevalence of .2 and below).

Figure \ref{fig:N10k_compare} shows relative performance, but it is also
important to consider how absolute uncertainty varies across conditions.
Figure \ref{fig:SE_plot} explores these relationships using
\citet{hanley1982meaning}'s formula for the approximate standard error
of the AUROC. The standard error is strongly decreasing in prevalence,
shown by the vertical separation between lines. The .01 condition has an
especially large standard error, making relative imprecision even more
costly in absolute terms. The standard error is mostly decreasing in
model quality, though interestingly, slightly increases from an AUROC of
0.5 before reaching a maximum around 0.6. As an aside, we can also form
``normal approximation'' confidence intervals by taking plus or minus
1.96 times the standard errors from Figure \ref{fig:SE_plot}. The normal
approximation intervals are typically close to the simulation confidence
intervals. For a prevalence = .01 and AUROC = .65 the simulation 95
percent confidence interval is {[}0.596, 0.702{]} while the normal
approximation is {[}0.591, 0.709{]}. The approximation becomes worse as
the AUROC increases: In the prevalence = .01 and AUROC = .95 condition
the simulation confidence interval is {[}0.929, 0.967{]} while the
normal approximation is {[}0.92, 0.98{]}. The adequacy of this
approximation bears on the utility of the alternative method for
estimating resolving power, described in Appendix A.

\begin{figure}
\includegraphics{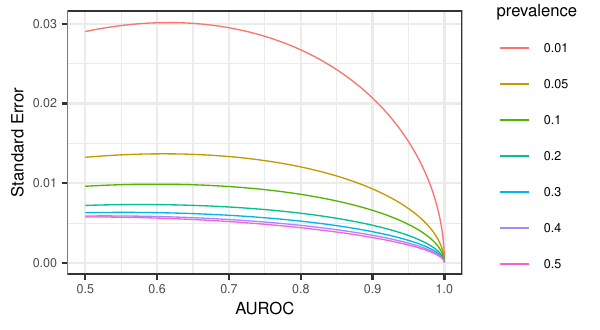} \caption{\label{fig:SE_plot}The relationship between AUROC and its standard error for different levels of outcome prevalence. The standard error is found using Hanley and McNeil (1982)'s approximation formula.}\label{fig:SE_plot}
\end{figure}

Thus, for moderately sized (N = 10,000) classification tasks the AUROC
will typically provide better resolution. Importantly, these results are
essentially unchanged for different sample sizes. For both a magnitude
smaller (N = 1000) and larger (N = 100,000), the AUROC is generally
better, with the AUPRC showing a relative advantage only among excellent
models with an outcome prevalence of 20 percent of less. Appendix B
presents results for these additional scenarios.

Now suppose the AUPRC is what we really want to maximize. That is, the
AUPRC best captures our intuitions about what makes a good model. But
also suppose that we know our resolving power assumptions are true: That
risk scores follow a binormal distribution and model improvements come
from additive shifts. Then what this section's results tell us is that
we should use the AUROC for search even if our goal is to find the best
AUPRC model (at least in most contexts). Under this hypothetical
scenario we don't need to balance the AUROC's greater resolving power
against our preference for the AUPRC because we know precisely when the
AUROC is a better guide. But once we select the final model we should
still describe its performance with the AUPRC since that is what we
really care about.

Moving from the hypothetical to reality, we know that this section's
assumptions will never fully hold. This means that the results should be
taken as only general guidance to be weighed against other criteria. For
instance, the AUPRC suffers only a modest disadvantage for ``good''
models with moderate prevalence. Once we factor the influence of
assumptions violations, which we could test with sensitivity analyses
that explore other paths towards class score separation, the
disadvantage may disappear.

It is uncertain how robust this section's guidance is to deviations from
the binormal model. One way to address this concern is to replace the
binormal with a domain-specific data-generating process. However, most
applications will not have the requisite quantitative framework to
accomplish this. An alternative strategy is to start with a dataset and
a baseline model and then build the sampling model from an initial set
of risk scores. This empirically-driven approach is explored in the next
section.

\section{Empirical sampling models}\label{sec6}

This section shows how specific problem information can be incorporated
into the resolving power approach. We will explore an example task where
the aim is to predict 30-day hospital readmissions among diabetes
patients using features such as patient demographics, prior utilization,
diagnoses, lab tests, and medications. The data for this example can be
found at the UCI Machine Learning Repository.\footnote{For access and a
  description of the original dataset go to
  \url{https://archive.ics.uci.edu/ml/datasets/diabetes+130-us+hospitals+for+years+1999-2008}.
  The post-processed data can be found at the GitHub address listed
  above.} After applying recommended restrictions, the dataset includes
69,973 total records with 6,277 readmissions for an outcome prevalence
of about 9 percent. So this is an example of an imbalanced class problem
for which the AUPRC is often recommended.

Now how can we use the data to guide our choice of a sampling model? A
seemingly sensible approach is to fit an initial classifier and then use
its risk scores to inform the choice. The initial model might be the
simplest algorithm among a set of candidates, or it could be a preferred
algorithm using its default hyperparameters. We may then construct a
sampling model from the empirical distribution (the set of outcomes and
risk scores) in a couple of different ways. One is to find a parametric
model that gives a good approximation to the empirical distribution.
Another is to treat the empirical distribution as the population, as is
done in resampling methods such as bootstrapping.

For the readmissions data, we use a simple logistic regression as the
initial model, estimating risk scores using 5-fold stratified
cross-validation. The estimated AUROC is .646 for this initial model. By
the previous section's taxonomy, this is a ``poor'' model with about a
10 percent outcome prevalence. So the binormal model results suggests
that we should prefer the AUROC to the AUPRC.

\begin{figure}
\includegraphics{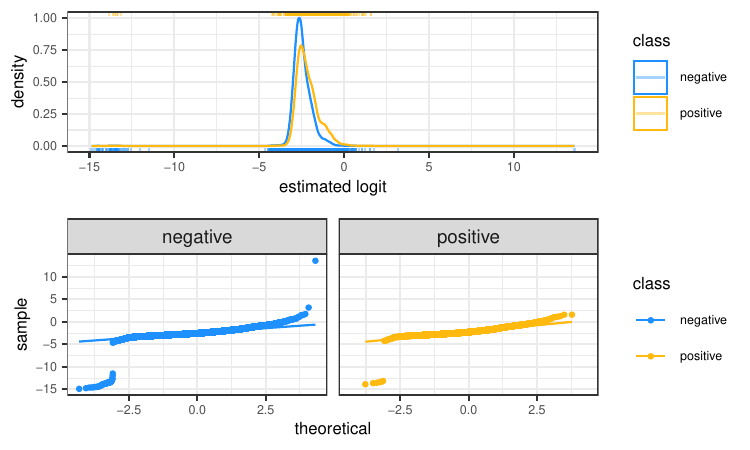} \caption{\label{fig:risk_distrib}Top panel: Estimated logit effects with kernel density estimates for the positive and negative class. Bottom panel: Q-Q norm plots of sample versus theoretical quantiles. Straight lines pass through the 1st and 3rd quartiles.}\label{fig:risk_distrib}
\end{figure}

The distribution of patient effects are shown in Figure
\ref{fig:risk_distrib}, with a rug plot and density estimates in the top
panel and qqnorm plots below. A normal approximation does not appear
appropriate as both the positive and negative class have large clusters
of scores in the lower tail. We could hunt for a better parametric
approximation---perhaps some type of mixture distribution---but instead
we will use the empirical distribution as the population sampling model.

Moving to the second step, how can we use an empirical sampling model to
construct a signal curve? A simple option is to add small increments to
all positive class scores, which is analogous to the approach we used
with the binormal model. Incremental improvement will then be
concentrated among the negative and positive cases with the closest risk
scores, wherever they may reside in the distribution. This seems
reasonable as it assumes that marginally better models will first amend
the ranking of cases that need the smallest adjustments.

The previous section created an evenly-spaced grid of AUROC values using
binormal model analytic results. An evenly-spaced grid is also possible
for an empirical distribution: To begin, suppose there are \(n^+\)
positive cases with risk scores \(r_i^+\) for \(i \in \{1, ..., n^+\}\)
and \(n^-\) negative cases with risk scores \(r_j^-\) for
\(j \in \{1, ..., n^-\}\). Further suppose all risk scores are unique
and that the classifier is not perfect, so \(r_i^+ < r_j^-\) for at
least one \((i,j)\) pair. This implies that
\(0.5 \leq \text{AUROC} < 1\) for the initial AUROC value. We will build
the grid in the direction of improving AUROC, though it is
straightforward to adapt the process for decreasing AUROC. First, find
\(\delta_1 = \text{min} \left( r^-_j - r^+_i |r^-_j>r^+_i\right)\), so
\(\delta_1\) is the smallest positive difference in risk scores between
two cases that are out-of-order such that the negative case is assigned
higher risk than the positive case. Similarly, we can find \(\delta_2\)
as the second smallest difference, \(\delta_3\) as the third smallest,
etc. Now if we add \(\delta_1 + \epsilon\) to all positive class risk
scores where \(\delta_1 < \delta_1 + \epsilon < \delta_2\) we will shift
the positive distribution just enough to resolve one pair of
out-of-order risk scores, but no more. From above we know that the AUROC
will then increase by \(\frac{1}{n^+}\times\frac{1}{n^-}\). If instead
we had added \(\delta_2 + \epsilon\) with
\(\delta_2 < \delta_2 + \epsilon < \delta_3\) then it would have fixed
two pairs of scores and the improvement would have been
\(\frac{2}{n^+ \times n^-}\). Thus, we can precisely increase or
decrease the AUROC in \(\frac{1}{n^+\times{n^-}}\) increments. The
result is useful for determining an initial increment to shift the class
scores. Achieving a fixed increment across the grid requires updating
the score distance calculations after each step, but this is
computationally costly and is typically unnecessary. Instead, it is most
important to choose an initial increment that creates a high density of
points across the grid range so that the signal curve may be reliably
estimated.

\begin{figure}
\includegraphics{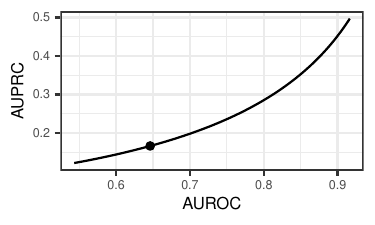} \caption{\label{fig:empirical_RC}An empirical signal curve from the readmissions data logistic regression model. The black point shows baseline performance. The curve is constructed by incrementing positive scores above and below the baseline distribution.}\label{fig:empirical_SC}
\end{figure}

Figure \ref{fig:empirical_RC} shows the signal curve constructed from
shifting the readmissions class score distributions. The starting model
AUROC and AUPRC values of .646 and 0.166 are shown by the dot. The curve
is built by shifting the positive class distribution above and below the
starting point, using an initial increment that produces a change of
.001 AUROC units. Each empirical distribution along the grid is
considered the population, so the associated population AUROC and AUPRC
are just the sample values. There are a total of 1000 grid points, which
range from .54 to .92 in AUROC and from .12 to .50 in AUPRC. In
practice, it is rare to see substantial improvement from initial
performance, so these ranges cover a larger space than is expected to be
observed during model search. The shape of the curve in Figure
\ref{fig:empirical_RC} is similar to the binormal signal curves: For
lower AUROC values the slope is relatively flat but then it increases
with improving model quality.

Moving to the third step of noise estimation, the initial model is the
natural choice to evaluate since improvements will be made from this
starting point. We generate 10,000 samples from the empirical
distribution, fixing the prevalence with stratified sampling, and then
find 95 percent confidence intervals. The fourth and final step uses the
signal curve to estimate and compare metric resolution. Results of the
sampling experiment are displayed in Table \ref{tab:empirical_summary}.

\begin{table}[!h]
\centering
\caption{\label{tab:empirical_summary}Simulation results summary. Lower and upper CI bounds are for 95 percent confidence intervals.}
\centering
\begin{tabular}[t]{>{}c|c|c|c|c}
\hline
\textbf{metric} & \textbf{Lower CI} & \textbf{Upper CI} & \textbf{$\kappa$} & \textbf{resolving power}\\
\hline
\textbf{AUROC} & 0.6391 & 0.6535 & 0.0144 & 69.4\\
\hline
\textbf{AUPRC} & 0.1601 & 0.1732 & NA & NA\\
\hline
\textbf{AUPRC to AUROC} & 0.6341 & 0.6591 & 0.0250 & 40.0\\
\hline
\end{tabular}
\end{table}

The last row uses the signal curve to map AUPRC to the AUROC scale.
Recall that metric resolution, \(\kappa\), is just the width of the 95
percent confidence interval in AUROC units. The AUROC has considerably
better discrimination with a resolving power that is over 70 percent
greater than the AUPRC. The absolute difference in confidence interval
widths is about 0.011 AUROC units, which could be substantial relative
to the often small improvements achieved during hyperparameter tuning.
Hence, the resolving power analysis predicts that the AUROC will be
better than the AUPRC for model search. But is it actually? The next
section considers the challenge of empirically assessing the resolving
power framework.

\section{Empirical validation}\label{sec7}

Resolving power's empirical importance crucially depends on the
topography of the model search space. Its impact will be limited when
search occurs in either ``signal dominant'' or ``noise dominant''
contexts. For instance, if the model evaluation points are spread across
a steep region of the space then the model quality signal will overwhelm
the noise for all metrics. In this case there will be consensus among
metrics so it will not matter which is used for model selection.
Conversely, some evaluation points may span an optimum that forms a flat
region of the search space. If the flat portion is the entire search
space or is sufficiently elevated (imagine an isolated butte) then again
the choice of evaluation metric will not matter. But instead of
unanimity we should now expect the metrics to agree only some of the
time.\footnote{If the metric estimates are independent across the
  optimum then the probability of consensus will be \((1/n)^{k-1}\)
  where \(n\) is the number of evaluation points on the optimum and
  \(k\) is the number of metrics.}

The above concerns noted, we still want to test the hypothesis that
higher resolving power leads to better model selection. An immediate
challenge is that for actual data, such as with the diabetes
readmissions example, we do not have access to the population. Yet with
a sufficiently large dataset we can approximate drawing samples from a
known population. This is the approach \citet{kohavi1995study} uses to
compare how different resampling methods affect classifier accuracy.
Since the population characteristics of real data cannot be known,
Kohavi uses the holdout method to estimate the ``true'' accuracies with
relatively large test sets.

This section employs a method similar to Kohavi's by repeatedly applying
the three-way-holdout method to the diabetes readmissions data. The
study uses 15 percent of the data for training, 5 percent for
validation, and 80 percent for the holdout set. The search space is over
a collection of XGBoost models, which is a popular implementation of
gradient boosted trees \citep{chen2016xgboost}. On each iteration
resolving power is first estimated using a baseline model with
\(\text{nrounds} = 25\) and a learning rate
\(\text{eta} = 0.1\).\footnote{See the online code supplement for the
  full model specification.} Model search then proceeds over six pairs
of randomly generated hyperparameter values with
\(\text{eta} \in [.01, .3]\) and \(\text{nrounds} \in [25, 150]\). The
experiment was repeated a total of 500 times with new data splits and
new hyperparameter values generated on each iteration. Since the goal is
to compare each metric's ability to guide search, both the AUROC and the
AUPRC are used for model selection on each iteration and then
performance is compared on the test set.

The AUROC had greater baseline estimated resolving power than the AUPRC
on 499 of the 500 trials, meaning that the AUROC would be preferred for
model search. For simplicity, the one aberrant trial is excluded from
the subsequent analysis. The two metrics agreed on the best model on 281
out of the remaining 499 trials. Of the 218 disagreements, using the
AUROC for model selection led to better AUROC test set performance on
146 of the trials and better AUPRC performance on 134 of the trials.
That is, it was better on 67 and 61 percent of trials, higher than the
predicted 50 percent if the two metrics were equally good for search.

\begin{table}[!h]
\centering
\caption{\label{tab:test_summary}The top row gives the average performance for trials where ROC and PRC model choice disagrees. Disagreement occurred on 218 out of 499 total trials. The bottom row gives performance on the trials where the two metrics agree.}
\centering
\begin{tabular}[t]{c|>{\centering\arraybackslash}p{2cm}|>{\centering\arraybackslash}p{2cm}|>{\centering\arraybackslash}p{2cm}|>{\centering\arraybackslash}p{2cm}}
\hline
\textbf{agree} & \textbf{tune ROC, test ROC} & \textbf{tune PRC, test ROC} & \textbf{tune ROC, test PRC} & \textbf{tune PRC, test PRC}\\
\hline
FALSE & 0.6205 & 0.6168 & 0.1495 & 0.1478\\
\hline
TRUE & 0.6227 & 0.6227 & 0.1509 & 0.1509\\
\hline
\end{tabular}
\end{table}

Table \ref{tab:test_summary} summarizes the average AUROC and AUPRC
scores when using the ROC versus PRC for model selection. Using the
AUROC gives an improvement of about .0037 AUROC units and about .0017
AUPRC units in average test performance (both significant at an
\(\alpha = .05\)). Thus, for this example we find evidence of an effect:
Greater resolving power leads to the discovery of better models as
scored by the ROC or the PRC curve.

For a given application the importance of resolving power will depend on
the classification task, the candidate evaluation metrics, and the set
of models being compared. Consequently, the observed effects of metric
choice, such as those from this section, are liable to shrink or grow
with any changes made to the model search space. This sensitivity to the
problem context poses a special challenge for assessing the general
empirical importance of resolving power.

\section{Conclusion}\label{sec8}

Evaluation metrics form the contours of a model's performance
topography, so choosing the right metric is essential for successful
navigation of this space. Resolving power is a framework for comparing
threshold-free metrics. Central to the method is the specification of a
class score sampling model that is used to both manipulate model quality
and probe sampling variability. The quality dimension, which serves as
the standard for comparison, is an ordered sequence of increasingly
separated score distributions. A pivotal assumption is that movement on
the quality dimension resembles the actual process of model improvement.
This paper uses simple additive shifts to separate the class scores.
Future work can test and refine this assumption by observing how risk
scores evolve as algorithms learn from real-world data. Signal curves
show how evaluation metrics respond to changes in classifier quality.
Metric error variance is found with random draws from the sampling
model. Resolving power is classifier-agnostic since it operates on risk
scores that are downstream of a classification model. Binormal model
simulation results provide general rules-of-thumb for when the AUROC
will have stronger resolving power than the AUPRC. The empirical method
allows researchers to use their data and an initial classifier of their
choice to construct a sampling model for estimating resolving power.

If we imagine evaluation metrics occupying a 2-dimensional space with
interpretability measured on the x-axis and resolving power on the
y-axis, then a metric's (intepretability, resolving power) coordinates
describe its usefulness for model search. A (low, low) metric has no
value. A (low, high) metric reliably maximizes an irrelevant or
inscrutable target. A (high, low) metric fails to maximize the desired
target. Of course (high, high) is best, but the challenge remains on how
to create these types of metrics.

\backmatter

\bmhead{Supplementary information}

Online Resource 1 is available at
\url{https://github.com/colinbeam/resolving_power/blob/main/S1_Online_Resource.pdf}

All code and simulation data used in the paper are available on GitHub
at \url{https://github.com/colinbeam/resolving_power}

\begin{appendices}

\section{Linear approximation method for resolving power}\label{secA1}

The resolving power approach from the main text has a couple of
disadvantages: It requires estimating the signal curves over many grid
points and choosing a specific \(\alpha\) value for the confidence
interval width. This section outlines a local, linear approximation that
avoids both of those drawbacks. Return to the toy example from Section
4, where we wish to compare the resolving power of the AUSGC versus the
AUROC. The linear approach is easier to express by flipping the axes,
plotting the AUSGC on the x-axis and the AUROC on the y-axis, as shown
in Figure \ref{fig:alt_resolution}.

Suppose we want to compare the resolving power of the AUSGC versus the
AUROC at the point (0.16, 0.8) shown in black. Figure
\ref{fig:alt_resolution} shows the full signal curve, but we only need
to evaluate a few points to estimate the tangent line shown in red. The
slope at the evaluation point, which gives the relative signal of the
two metrics, is 0.69. Since the slope is less than 1 it means the AUSGC
has a relatively stronger response to improvements in model quality.

Next, we must estimate sampling variability at the evaluation point.
Again, suppose we form many simulation samples of the evaluation
metrics. But instead of finding percentile confidence intervals, we use
the simulation samples to estimate the standard deviation of the AUSGC
and AUROC, denoted respectively as \(\hat{\sigma}_{\text{S}}\) and
\(\hat{\sigma}_{\text{R}}\).\footnote{The standard deviation of the
  simulation samples estimates the standard error of the evaluation
  metric.} If we assume that the distribution of the sample AUROC is
approximately normal, then we can form a \(1 - \alpha\) confidence
interval by selecting a \(z_{(1-\alpha/2)}\) critical value and
multiplying the standard error. Using
\(z_{(1-\alpha/2)}=1.96 \approx 2\) gives an approximate 95 percent
confidence interval. So for the AUROC we get the interval:
\([-2\hat{\sigma}_{\text{R}}, \; 2\hat{\sigma}_{\text{R}}]\). The metric
resolution (the width of the confidence interval in AUROC units) is then
\(\kappa_{\text{ROC}} = 4\hat{\sigma}_{\text{R}}\). Similarly, the
approximate 95 percent confidence interval for the AUSGC is
\([-2\hat{\sigma}_{\text{S}}, \; 2\hat{\sigma}_{\text{S}}]\). Now we use
the linear approximation to map AUSGC to the AUROC scale. Suppose the
slope of the linear approximation is \(\beta_1\) and the intercept is
\(\beta_0\), then we obtain the confidence interval
\([-2\hat{\sigma}_{\text{S}}\beta_1 + \beta_0, \; 2\hat{\sigma}_{\text{S}}\beta_1+\beta_0]\)
and its width is
\(\kappa_{\text{SGC}} = 4\hat{\sigma}_{\text{S}}\beta_1\). Taking the
ratio of metric resolutions gives:

\[ \frac{\kappa_{\text{SGC}}}{\kappa_{\text{ROC}}} = \frac{4\hat{\sigma}_{\text{S}}\beta_1}{ 4\hat{\sigma}_{\text{R}}} = \frac{\beta_1\hat{\sigma}_{\text{S}}}{ \hat{\sigma}_{\text{R}}} \]
The \(z\) critical value cancels and we are left with the ratio of
standard errors scaled in AUROC units. Thus, the linear approach
requires only comparing the ratio of the standard errors to the slope of
the signal curve, eliminating the need for an \(\alpha\) level. Using
the Leibniz notation \(\beta_1 = \frac{dR}{dS}\), we must only check the
inequality: \[
\tag{A.1}
\frac{dR}{dS} > \frac{\hat{\sigma}_{\text{R}}}{\hat{\sigma}_{\text{S}}} 
\] Inequality A.1 makes transparent the signal to noise comparison: The
left side is the relative signal of the two metrics while the right side
is the relative noise. If the inequality holds it means that the signal
of the AUROC overwhelms its noise, giving it relatively greater
resolving power.

\begin{figure}
\includegraphics{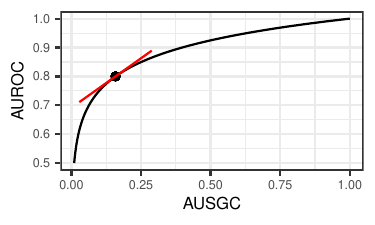} \caption{\label{fig:alt_resolution}Signal curve plotting the AUROC reference scale on the y-axis. A linear approximation of the signal function at the evaluation point is shown in red.}\label{fig:alt_resolution}
\end{figure}

The simplicity of the linear approach bears both its strengths and
weaknesses. Estimating the tangent curve requires only a few points in
the immediate vicinity of the evaluation point, obviating the need to
find the full signal curve or choose a confidence interval width. The
approach crucially assumes:

\begin{enumerate}
\def\labelenumi{\roman{enumi}.}
\tightlist
\item
  The metric sampling distribution is symmetric.
\item
  A line is a good approximation of the signal curve over the region of
  interest.
\end{enumerate}

Recall that we calculate metric resolution by using the signal curve to
map confidence interval limits from one scale to another. A linear
function will give a satisfactory approximation when these confidence
limits are narrow. However, it will be poor for wide confidence
intervals bracketing a curve segment that has a rapidly changing slope.

\section{Additional binormal results}\label{secA2}

\begin{figure}
\includegraphics{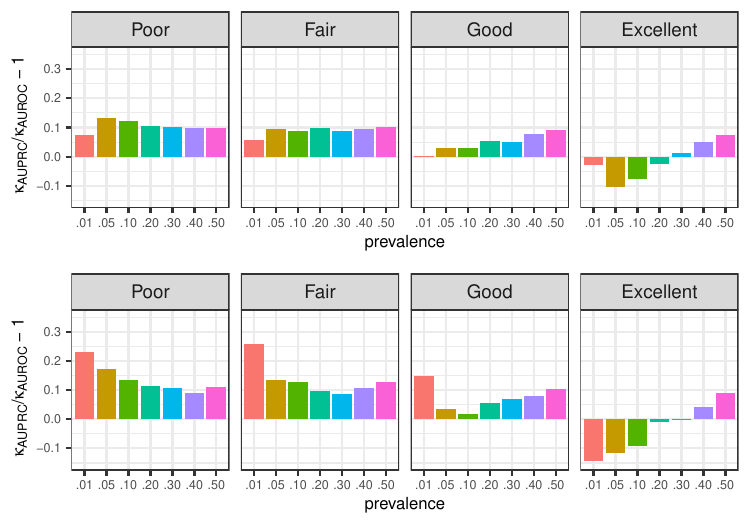} \caption{\label{fig:additional_binormal}Relative metric resolution by outcome prevalence and model quality for a binormal model with a sample size of top panel: N = 1000 and bottom panel: N = 100,000. At each level of model quality 10,000 simulations are taken from the sampling model. Confidence limits are the .025 and .975 quantile values of the simulation samples.}\label{fig:additional_binormal}
\end{figure}

We extend the binormal investigation to sample sizes that are an order
of magnitude smaller and larger than those in the main text, shown
respectively in the top and bottom panels of Figure
\ref{fig:additional_binormal}. For the \(N = 1000\) study the simulation
was again repeated three times with estimates averaged across runs. In
the \(N = 100,000\) study the simulation was conducted only once.

Interestingly, for \(N=1000\) the direction of the differences across
conditions are the same as found in the \(N = 10,000\) study. Only the
relative magnitudes have changed. Specifically, the AUPRC has superior
resolving power only for ``excellent'' models with an outcome prevalence
of 20 percent or less. The primary difference in magnitudes are found in
the one percent prevalence condition, which now shows a smaller
disadvantage for the fair to good models, and a smaller advantage for
the excellent models. Note that there are now only ten instances in the
positive class for one percent prevalence.

Results from the increased order of magnitude \(N = 100,000\) study are
also similar. There is one condition where the direction of the effect
has flipped---the excellent model condition with a prevalence of 30
percent---though the relative difference is essentially zero. The other
primary difference is that the AUPRC is now at the biggest disadvantage
for the one percent prevalence ``fair'' model condition.

In summary, differences in sample size for the binormal model generally
do not affect the direction of the effects, only their relative size.
The AUPRC maintains an advantage only for excellent models with an
outcome prevalence of 20 percent or less.

\end{appendices}

\bibliography{references.bib}

\end{document}